\documentclass[a4paper,superscriptaddress,nofootinbib,12pt]{article}
\usepackage{amssymb}
\usepackage{eurosym}
\usepackage{amsfonts}
\usepackage{geometry}
\usepackage{color}
\usepackage[normalem]{ulem}
\begin{document}

\date{}

\vspace{1cm}

\begin{center}
\large{\textbf{Generalized quantum phase spaces for the $\kappa$-deformed extended Snyder model}} \\[0pt]

\vskip1.0cm
\normalsize

\textbf{Jerzy Lukierski$^{1}$, Stjepan Meljanac$^{2}$, Salvatore Mignemi$^3$, Anna Pacho\l$^{4}$} \\[0pt]
\vskip0.5cm
{\small{\ {$^{1}$}}Institute of Theoretical Physics, Wroc\l aw University, pl. Maxa Borna 9, 50-205 Wroc\l aw, Poland}

 {\small {\ {$^{2}$ }}Division of Theoretical Physics, Rudj{}er Bo\v{s}kovi\'c Institute, Bijeni\v{c}ka~c.54, HR-10002~Zagreb, Croatia}

{\small $^{3}$
Dipartimento di Matematica, Universit\`a di Cagliari
via Ospedale 72, 09124 Cagliari, Italy
and INFN, Sezione di Cagliari}

{\small $^{4}$
Department of Microsystems, University of South-Eastern Norway, Campus Vestfold, Raveien 215, 3184 Borre, Norway}
\end{center}

\vspace{1cm}
\begin{abstract}
We describe, in an algebraic way, the  $\kappa$-deformed extended Snyder models,
 that depend on three parameters $\beta, \kappa$ and $\lambda$,
   which in a suitable algebra basis are described by the de Sitter algebras
    ${o}({1,N})$.
     The commutation relations of the algebra contain a parameter $\lambda$,
   which is used for the calculations of perturbative expansions.
   For such $\kappa$-deformed extended
  Snyder models we consider   the
Heisenberg double with  dual generalized momenta sector, and
  provide  the  respective
   generalized quantum phase space  depending on three parameters
    mentioned above.
   Further, we study for these models
   an alternative Heisenberg double, 
  with the algebra of functions on de Sitter group.
      In      both cases     we   calculate the formulae
   for the cross commutation relations   between generalized coordinate and momenta sectors,
   at linear order in $\lambda$.
     We  demonstrate that in the commutators of quantum  space-time coordinates and momenta
of the quantum-deformed Heisenberg algebra  the terms
 generated by  $\kappa$-deformation are dominating over $\beta$-dependent ones
 for small values of $\lambda$.
\end{abstract}

\section{Introduction}

The non-commutative (NC)
 quantum space-times, where the coordinates are elements of NC  quantum space-time algebra,
  have been considered as a  tool for the
  description of quantum gravity  (QG)   (see, e.g. \cite{Dop1}, \cite{S. Majid}), which unifies two
basic theories in physics: general relativity and quantum mechanics. One
of the first examples of quantum space-time was  proposed by Snyder  already in 1947 \cite{Snyder40}.
   Other  quantum space-time models were studied in the 1990's   (e.g. $\kappa$-deformed relativistic theories    \cite{LuNoRu}-\cite{PodWo}).
  Due to the increasing  interest in QG, the framework of quantum symmetries described by
 the theory of quantum groups was developed.
 The NC   structure of  space-time at the Planck scale appeared as a way
 to describe QG  at very short distances and led to the description of quantum-mechanical and quantum-deformed relativistic phase spaces.
 It appeared that these  NC structures of quantum space-times and quantum
 relativistic {symmetries} are associated
 with the formalism of Hopf algebras and the case of  quantum phase spaces has been generalized
 to the Hopf algebroids framework (see, e.g.~\cite{JHL}, \cite{PXu}).

In this paper we find the quantum-deformed relativistic phase space for the
$\kappa$-deformed extended Snyder model \cite{MM_2021}\footnote{We use ``$\kappa$-extended Snyder model''   as a short notation for
the $\lambda$-dependent $\kappa$-deformed extended Snyder model.} which will be  obtained by  the Heisenberg double construction.
   The  Heisenberg  double,  which constitutes a Hopf algebroid, is
     defined  within the Hopf algebraic framework and can be considered as  providing the way to
 define quantum-deformed phase spaces in the presence of QG effects.

The $\kappa$-{deformed} Snyder model,
proposed firstly in \cite{MMSS}, \cite{AMDMASMS},
 unifies the two perhaps  best known models which   describe the NC
  relativistic space-time coordinates $\hat{x}_i$:\footnote{N-dimensional Latin indices include zero,
  which  describes time-like dimension, i.e. $i,j=0,\ldots, N-1$.}
 \newline
i) the Snyder model:
\begin{equation}
[\hat{x}_i,\hat{x}_j]=i\beta M_{ij}
\end{equation}
where $\beta$ is a parameter with dimension of length square and $M_{ij}$
are the generators of Lorentz transformations;\newline
ii)  $\kappa $-Minkowski quantum space-time:

\begin{equation}\label{km}
\lbrack \hat{x}_{i},\hat{x}_{j}]=i(a_{i}\hat{x}_{j}-a_{j}\hat{x}_{i})
\end{equation}%
where $a_{i}$ is a  constant  vector
multiplied by the parameter $\kappa^{-1}$ with dimension of length.
Both Snyder and $\kappa$-deformed Snyder models lead to non-associativity and non-coassociativity \cite{Battisti:2010sr}, \cite{Girelli}. In order to avoid problems with non-associativity, an  alternative Snyder model, in which {Snyder space} is a subspace of a larger non-commutative space was proposed in \cite{Girelli} for the 3 dimensional Euclidean Snyder space, and its generalization to the extended Snyder model was proposed in \cite{2007Snyder}
\footnote{The $N$-dimensional extended Snyder algebra \cite{Girelli}, after introducing the $\lambda$-dependence 
\cite{2007Snyder}, takes the form of an $N$-dimensional de Sitter algebra given by  ($\mu=0,1\ldots N$)
\begin{equation}\label{3footnote}
\lbrack \hat{x}_{\mu \nu },\hat{x}_{\rho \sigma }]=i\lambda (\eta _{\mu \rho
}\hat{x}_{\nu \sigma }-\eta _{\nu \rho }\hat{x}_{\mu \sigma }+\eta _{\nu
\sigma }\hat{x}_{\mu \rho }-\eta _{\mu \sigma }\hat{x}_{\nu \rho })
\end{equation}
where
$\hat{x}_{\mu \nu }$ are the standard ${o}(1{,}N)$ generators. 
A specific new feature of such  formulation of  extended Snyder algebra introduced in \cite{2007Snyder}, is the appearance of a dimensionless parameter $\lambda$ such that for $\lambda\rightarrow 0$ the extended Snyder algebra reduces to the Abelian algebra { and generators $\hat{x}_{\mu \nu }$ become commutative}.
The extended Snyder algebra (\ref{3footnote})
contains the generators defining  Snyder quantum space-time coordinates $\hat{x}_i$ ({for which the classical limit is obtained when $\beta\rightarrow 0$}) as well as tensorial coordinates $\hat{x}_{ij}$   (see (\ref{Snyder},\ref{kappa})). {For the interpretation of $\hat x_{ij}$ we refer the reader to \cite{LMMP_12_2022}}.}. If  one uses {Latin} indices
 the algebra describing the
  $\lambda$-dependent extended Snyder model  
  looks as follows  \cite{2007Snyder}
\begin{eqnarray}
\lbrack \hat{x}_{i},\hat{x}_{j}] &=&i\lambda \beta \hat{x}_{ij},\qquad
\lbrack \hat{x}_{ij},\hat{x}_{kl}]=i\lambda (\eta _{ik}\hat{x}_{jl}-\eta
_{il}\hat{x}_{jk}-\eta _{jk}\hat{x}_{il}+\eta _{jl}\hat{x}_{ik}),
\label{Snyder} \\
\lbrack \hat{x}_{ij},\hat{x}_{k}] &=&i\lambda (\eta _{ik}\hat{x}_{j}-\eta
_{jk}\hat{x}_{i}),  \label{kappa}
\end{eqnarray}%
where $\hat{x}_i = \sqrt{\beta}  \hat{x}_{iN}$.

{Unification of $\kappa$-Minkowski and extended Snyder space-times was proposed in \cite{MM_2021} generalizing the extended Snyder model \cite{2007Snyder} by including the $\kappa$-Minkowski algebra terms (2) using orthogonal algebra with metric tensor $g$ \cite{BorPa}, \cite{ABAP-EPJC}.  The main step for achieving this unification was introducing particular modification of constant metric tensor instead of Minkowski metric.}

In this paper we introduce
the  generalization of quantum-mechanical phase space {corresponding to}
  the $\kappa$-extended Snyder
   model.
   The phase spaces   with
      NC space-time coordinates, have been already considered
        for       Snyder model and for models covariant under  the
  $\kappa$-deformed Poincar\'e algebra:  the generalized phase spaces
   containing the $\kappa$-Minkowski NC
space-time
 were considered in
   \cite{KosMas} -
   \cite{Meljanac:2021qgq}
  and
 the phase space for the Snyder model in, e.g.  \cite{Snyder40}, \cite{Battisti:2008xy}, \cite{Mignemi}
 as well as for the extended Snyder model in  \cite{MP_2021}. 
     Quantum space-times and deformed phase spaces
 for  Snyder and Yang models were  discussed in
\cite{Lukierski:2021tam},  \cite{LukWor} 
 together with their extensions to supersymmetric models, see also \cite{Meljanac:2022qhp}.

Here we propose the description of the corresponding deformed
 phase spaces for the $\kappa$-extended Snyder model. The unification of the extended Snyder and $\kappa$-Minkowski models
 proposed in
 \cite{MM_2021}, \cite{MM_2021Sept} 
 is  realized   in the framework of associative and   coassociative Hopf algebra,
   what permits to apply the Heisenberg double construction
   that   is  considered in the present  paper. The resulting generalized phase spaces offer interesting
    insights into the $\kappa$-extended Snyder model,
  which is described by the superposition of two quantum deformations
   with a further parameter $\lambda$.

\section{$\kappa$-deformed extended Snyder model}

Let us recall the algebra corresponding to the $\kappa$-extended Snyder model, which was introduced in \cite{MM_2021} as the extended unified $\kappa$-Minkowski Snyder model.  We shall denote it as $o(1,N;g)$.
It is defined by the following
 Lie-algebraic set of  commutation relations:
\begin{equation}
\lbrack \hat{X}_{\mu \nu },\hat{X}_{\rho \sigma }]=i\lambda (g_{\mu \rho }%
\hat{X}_{\nu \sigma }-g_{\nu \rho }\hat{X}_{\mu \sigma }+g_{\nu \sigma }\hat{%
X}_{\mu \rho }-g_{\mu \sigma }\hat{X}_{\nu \rho }).  \label{alg_sog}
\end{equation}
The metric $g \equiv  (g_{\mu \nu })$ has the form:
\begin{center}
$g=\left(
\begin{array}{ccccc}
-1 & 0 & ... & 0 & g_{0} \\
0 & 1 & ... & 0 & g_{1} \\
.. & .. & .. & .. & ... \\
0 & 0 & .. & 1 & g_{N-1} \\
g_{0} & g_{1} & ... & g_{N-1} & g_{N}%
\end{array}%
\right) $
\end{center}
with $\det g=-g_{0}^{2}+\sum_{i=1}^{N-1}g_{i}^{2}-g_{N}$. If we rewrite the
 generators
 $\hat{X}_{\mu \nu }$ as $\hat{X}_{ij}$ and $\hat{X}_{kN}=\kappa \hat{X}_{k}$
 where $\kappa $ is a new mass-like parameter, and rewrite the metric as
follows
 $$g_{ij}=\eta _{ij},\quad g_{iN}=\kappa a_{i},\quad
g_{NN}=g_{N}=\kappa ^{2}\beta $$
 the algebra  (\ref{alg_sog})
splits  to the following set of relations
\begin{eqnarray}
\lbrack \hat{X}_{i},\hat{X}_{j}] &=&i\lambda (a_{i}\hat{X}_{j}-a_{j}\hat{X}%
_{i}+\beta \hat{X}_{ij}), \\
\lbrack \hat{X}_{ij},\hat{X}_{k}] &=&i\lambda (\eta _{ik}\hat{X}_{j}-\eta
_{jk}\hat{X}_{i}+a_{j}\hat{X}_{ik}-a_{i}\hat{X}_{jk}), \\
\lbrack \hat{X}_{ij},\hat{X}_{kl}] &=&i\lambda (\eta _{ik}\hat{X}_{jl}-\eta
_{jk}\hat{X}_{il}+\eta _{jl}\hat{X}_{ik}-\eta _{il}\hat{X}_{jk}).
\end{eqnarray}%
From the first commutator we see how the Snyder and $\kappa $-Minkowski
space-time relations are unified. The $\hat{X}_{i}$ are the NC space-time
coordinates, and $\hat{X}_{ij}$ can be interpreted as non-commutative
tensorial coordinates.

Moreover, when $g_{N}=0$,  the relations (\ref{alg_sog})
   describe the  $\kappa $-Minkowski space-time with Lorentz covariance algebra.
   Alternatively,  when
$g_{0}=...=g_{N-1}=0$ and $g_{N}=1$, then the relations (\ref{alg_sog}) reduce
to the algebra describing the extended  Snyder model.
   Also note that
if $g_{\mu \nu }\rightarrow \eta _{\mu \nu }$ then the algebra
 (\ref{alg_sog}) becomes a standard orthogonal algebra,
  which describes the $N$-dimensional  suitably rescaled de Sitter
   algebra.

The coalgebra sector is a classical one (primitive), i.e.
\begin{equation}
\Delta \left( \hat{X}_{\mu \nu }\right) =\Delta _{0}\left( \hat{X}_{\mu \nu
}\right) ,  \label{copx}
\end{equation}%
\begin{equation}
\epsilon \left( \hat{X}_{\mu \nu }\right) =0\mbox{\ and\ }S\left( \hat{X}%
_{\mu \nu }\right) =-\hat{X}_{\mu \nu } \label{Sx}
\end{equation}%
and reduces accordingly to coproducts, counits and antipodes for $\hat{X}_{k}$ and $\hat{%
X}_{ik}$.

If one defines the following change of coordinates:
\begin{equation}
\hat{X}_{\mu \nu }=\left( O\hat{x}O^{T}\right) _{\mu \nu }~,\quad g_{\mu \nu
}=\left( O\eta O^{T}\right) _{\mu \nu }~,
\end{equation}
with the  
 choice of the matrix
\begin{equation}
O=\left(
\begin{array}{ccccc}
1 & 0 & ... & 0 & 0 \\
0 & 1 & ... & 0 & 0 \\
.. & .. & .. & .. & ... \\
0 & 0 & .. & 1 & 0 \\
-g_{0} & g_{1} & ... & g_{N-1} & \rho%
\end{array}%
\label{macierz}
\right)
\end{equation}
where $\rho =\sqrt{g_{N}-g_{k}g_{k}}=\sqrt{-\det g}$ , one can show that the
algebra (\ref{alg_sog}) reduces to (see also ${}^{3)}$)
\begin{equation}
\lbrack \hat{x}_{\mu \nu },\hat{x}_{\rho \sigma }]=i\lambda (\eta _{\mu \rho
}\hat{x}_{\nu \sigma }-\eta _{\nu \rho }\hat{x}_{\mu \sigma }+\eta _{\nu
\sigma }\hat{x}_{\mu \rho }-\eta _{\mu \sigma }\hat{x}_{\nu \rho }),
\label{extSnyder}
\end{equation}
where $\eta _{\mu \nu }=\left( -1,1,...,1\right) $ and $g_{N}=1$. This algebra has been discussed in \cite{Girelli}, \cite{2007Snyder}, as the
extended Snyder algebra.   
 We recall (see ${}^{3)}$ and (4), (5)) that the
 algebra  (14) reduces to the
   $N$-dimensional Snyder space extended by the Lorentz algebra.
   We add that
  Heisenberg double for the
extended Snyder model (\ref{extSnyder}) was studied in \cite{MP_2021}.

Due to the form of the matrix $O$ we have the following relations between
two sets of coordinates:
\begin{equation}  \label{Xxrel}
\hat{X}_{i}=\rho \hat{x}_{i}+a_{j}\hat{x}_{ij},\qquad \hat{X}_{ij}=\hat{x}%
_{ij}.
\end{equation}
Both sets of the coordinates $\hat{X}_{\mu \nu }$ and $\hat{x}_{\mu \nu }$
describe NC extended space-time and are related via relations (\ref{Xxrel}).

In the classical limit ($\lambda \rightarrow 0$) the generators of the
algebras (\ref{alg_sog}) and (\ref{extSnyder})
  reduce to the commutative  (Abelian) ones \cite{MM_2021Sept}
  $\hat{X}_{\mu \nu
}\rightarrow X_{\mu \nu }$ with $[X_{\mu \nu },X_{\rho \sigma }]=0$; ($\hat{x}_{\mu \nu }\rightarrow x_{\mu \nu }$ with $[x_{\mu \nu },x_{\rho \sigma }]=0$).
 These Abelian coordinates are
related with each other via the matrix $O$ given by
                                 (\ref{macierz})     
\[
X_{\mu \nu }=\left( OxO^{T}\right) _{\mu \nu },
\]
i.e. explicitly:
\[
X_{i}=\rho x_{i}+a_{j}x_{ij},\qquad X_{ij}=x_{ij},
\]
where 
  $X_{i}=\frac{1}{\kappa }X_{iN}$ and $x_{i}=\frac{1}{\kappa
}x_{iN}$.
 One can say that the algebra (\ref{alg_sog}),
   with the generators $\hat{X}%
_{\mu \nu }$ can be seen as the deformation of the underlying commutative space described
by $X_{\mu \nu }$, with
$\lambda $ as the deformation parameter.

For the  commutative extended spacetime coordinates $X_{\mu \nu }$ and $x_{\mu
\nu }$ we can introduce the extended momenta
($P^{\mu\nu}, p^{\mu\nu}$)
which
can be realized in a standard way as $P^{\mu \nu }=-i\frac{\partial }{\partial
X_{\mu \nu }}$ and $p^{\mu \nu }=-i\frac{\partial }{\partial x_{\mu \nu }}$.
 In this way  one can introduce
 two copies of the generalized Heisenberg algebra
 as unital, associative algebras generated by the extended
coordinates and momenta, with the tensorial coordinates  antisymmetric
under the exchange $\mu \leftrightarrow \nu$.
The following commutation relations are valid (we put $\hbar=1$)
\begin{equation}
\left[ X_{\mu \nu },X_{\alpha \beta }\right] =0=[P^{\mu \nu },P^{\alpha
\beta }],\qquad \left[ X_{\mu \nu },P^{\rho \sigma }\right] =i\left( \delta
_{\mu }^{\rho }\delta _{\nu }^{\sigma }-\delta _{\mu }^{\sigma }\delta _{\nu
}^{\rho }\right) ,  \label{PX}
\end{equation}
and similarly
\begin{equation}
\left[ x_{\mu \nu },x_{\alpha \beta }\right] =0=[p^{\mu \nu },p^{\alpha
\beta }],\qquad \left[ x_{\mu \nu },p^{\rho \sigma }\right] =i\left( \delta
_{\mu }^{\rho }\delta _{\nu }^{\sigma }-\delta _{\mu }^{\sigma }\delta _{\nu
}^{\rho }\right) ,  \label{px}
\end{equation}
where Greek indices are raised and lowered by the metric $g_{\mu \nu }$.

One can introduce, as well, the  formulae for the momenta with a single index
  (i.e  $P^{i}=\kappa P^{iN},p^{i}=\kappa p^{iN}$)
 as dual to space-time coordinates $X_i$ and $x_i$;
Latin indices are raised and lowered by the
flat metric $\eta _{ij}$. We also have analogous relations between the two types
of momenta (\ref{PX}) and (\ref{px})
\begin{equation}\label{Pprel}
p_{i}=\rho P_{i},\qquad p_{ij}=P_{ij}-a_{i}P_{j}+a_{j}P_{i},
\end{equation}
with the relation $P^{\mu \nu }=\left( \left( O^{-1}\right) ^{T}p\left(
O^{-1}\right) \right) ^{\mu \nu }$.
For more details about  these coordinates and momenta
 we refer the reader to \cite{MM_2021}, \cite%
{MM_2021Sept} where this model was introduced and studied.

In order to discuss  the phase spaces associated with the $\kappa $%
-deformed extended Snyder model (\ref{alg_sog}) one can
 use the
Heisenberg double construction method.

\subsection{Commutative momenta for the $\kappa$-extended Snyder model
 and their coproducts}

One can introduce the  Abelian momenta, dual %
 to the coordinates $\hat{X}_{\mu \nu }$ which describe the algebra $o(1,N;g)$
  (see (\ref{alg_sog})),
  by deforming
the canonically conjugate Abelian  momenta $P^{\mu \nu }$.
The technique to calculate the coproducts
 $\Delta P_{i}$ and $\Delta P_{ij}$ {was} proposed
  in \cite{AMDMASMS}, \cite{Battisti:2010sr}.
The commutation
relations for $P^{i}=\kappa P^{iN},\ P^{ij}$ remain  unchanged:
\begin{equation}
\left[ P_{i},P_{j}\right] =0,\qquad \left[ P_{ik},P_{jl}\right] =0,\qquad %
\left[ P_{i},P_{jl}\right] =0.  \label{PP}
\end{equation}%
 while the coalgebraic sector  of the   momentum generators
 \cite{MM_2021Sept} looks as follows\footnote{
Note that there was a typo in sign in term $P_{jk}\otimes P_i a_k$ in the coproduct of $\Delta P_{ij}$ in \cite{MM_2021Sept}.}:
\begin{eqnarray}
\Delta P_{i} &=&P_{i}\otimes 1+1\otimes P_{i} \nonumber \\
&&+\lambda \lbrack -c_{1}(P_{j}\otimes P_{ij}+P_{j}\otimes P_{i}a_{j})\nonumber \\
&&+(1-c_{1})(P_{ij}\otimes P_{j}+P_{i}\otimes
P_{j}a_{j})+(2c_{1}-1)P_{j}\otimes P_{j}a_{i}]+O(\lambda ^{2})\nonumber\\
&&
\label{cop_p1} \\
\Delta P_{ij} &=&P_{ij}\otimes 1+1\otimes P_{ij}  \nonumber \\
&&+\frac{\lambda }{2}[-\beta P_{i}\otimes P_{j}-(P_{ik}\otimes
P_{jk}+a_{k}\left( P_{i}\otimes P_{jk}-P_{jk}\otimes P_{i}\right) )\nonumber \\
&& -(2c_{1}-1)
(P_{k}\otimes P_{jk}+P_{k}\otimes P_{j}a_{k}+P_{jk}\otimes
P_{k}+P_{j}\otimes P_{k}a_{k})a_{i}-(i\leftrightarrow j)]+O\left( \lambda
^{2}\right) , \nonumber \\
&& \label{cop_p2} \\
S\left( P_{i}\right)  &=&-P_{i}+\lambda \left( 1-2c_{1}\right) \left(
P_{j}P_{ij}+a_{k}P_{k}P_{i}-P_{k}P_{k}a_{i}\right) , \\
S\left( P_{ij}\right)  &=&-P_{ij}-\lambda \left( 2c_{1}-1\right) \left(
a_{i}P_{jk}P_{k}+a_{i}P_{j}a_{k}P_{k}-(i\leftrightarrow j)\right),\\
\epsilon \left( P_{i}\right)  &=&\epsilon \left( P_{ij}\right) =0.\label{SP}
\end{eqnarray}
where a new parameter $c_{1}$
 depends on the realization
   (see Sec. 3 in \cite{MM_2021Sept}), and coproducts  are presented
     up to
the first order in the deformation parameter $\lambda $.
 The  coproducts (\ref{cop_p1}), (\ref{cop_p2})
correspond to the so-called generic realization \cite{MM_2021Sept}, and define the
momentum sector of the $\kappa$-extended Snyder model as parametrized by
  $\lambda, \beta, a_i $ and $c_{1}$.
  Such parametrization  occurs
 when the  coproducts  are written up to the first
order in $\lambda $; in higher orders of $\lambda $ 
 additional  parameters appear
 (see Sec.3 in \cite{MM_2021Sept} for more details).
 For the concrete Weyl
realization \cite{MM_2021}, one should set  $c_{1}=
\frac{1}{2}$ and
for the natural realization (i.e.  with  classical algebra basis) $c_{1}=0$.

We postulate the   standard  duality  relation:
\begin{eqnarray}
&<&P_{j},\hat{X}_{i}>=-i\eta _{ij}, \label{dual1} \\
&<&P_{k},\hat{X}_{ij}>=0, \\
&<&P_{kl},\hat{X}_{i}>=0, \\
&<&P_{kl},\hat{X}_{ij}>=-i\left( \eta _{ik}\eta _{jl}-\eta _{jk}\eta
_{il}\right) .\label{dual2}
\end{eqnarray}
One can check that from  the duality relation between the products
  in the algebra generated by $\hat{X}$ and the coproducts
in the coalgebra generated by $P$,
 the  following
   compatibility conditions hold
\begin{eqnarray}
&<&b_{\left( 1\right) },a><b_{\left( 2\right) },a^{\prime }>=<b,a\cdot
a^{\prime }>,
\\
&<&b,a_{\left( 1\right) }><b^{\prime },a_{\left( 2\right) }>=<b\cdot
b^{\prime },a>. 
\end{eqnarray}
\section{Generalized quantum phase space from the Heisenberg double}

We introduce the left Hopf action $\triangleright $
of momenta on coordinates  defined by the formula\\
$
P\triangleright \hat{X}=<P,\hat{X}_{\left( 2\right) }>\hat{X}_{{\left( 1\right) }}
$,
 and recall that  we use the Sweedler notation for the coproduct. From (\ref{dual1})-(\ref{dual2})
   it follows immediately that
$P_{j}\triangleright\hat{X}_{i}=-i\eta _{ij}, \quad P_{k}\triangleright\hat{X}_{ij}=0, \quad
P_{kl}\triangleright\hat{X}_{i}=0,$ \\
$P_{kl}\triangleright\hat{X}_{ij}=-i\left( \eta _{ik}\eta _{jl}-\eta _{jk}\eta
_{il}\right).$
The corresponding Heisenberg double commutators follow from the cross product
construction:
\begin{equation}
\left[ P,\hat{X}\right] =\hat{X}_{\left( 1\right) }<P_{\left( 1\right) },\hat{X}_{
\left( 2\right) }>P_{\left( 2\right) }-\hat{X}P.
\end{equation}
written shortly without the indices.
Doing the calculation explicitly and using the coproducts for momenta (\ref%
{cop_p1})-(\ref{cop_p2}), case by case, we obtain:
\begin{eqnarray}
\left[ P_{j},\hat{X}_{i}\right] &=&-i\eta _{j}{}_{i}+i\lambda \lbrack
c_{1}\left( P_{ji}+P_{j}a_{i}\right) -\left( 1-c_{1}\right) \eta
_{ij}P_{r}a_{r}-\left( 2c_{1}-1\right) P_{i}a_{j}]+O\left( \lambda
^{2}\right) , \label{pjxi} \\
\left[ P_{j},\hat{X}_{is}\right] &=&-i\lambda \left( 1-c_{1}\right) \left(
\eta _{ij}P_{s}-\eta _{sj}P_{i}\right) +O\left( \lambda ^{2}\right),
\label{pjxis}
\end{eqnarray}
\begin{eqnarray}
\left[ P_{ij},\hat{X}_{k}\right] &=&\frac{i\lambda }{2}\{\beta (\eta_{i}{}_{k}P_{j}-\eta_{jk}P_i)
+(\eta_{ik}P_{jl}-\eta_{jk}P_{il})a_{l}\nonumber\\
&&+\left(2c_{1}-1\right) [a_{i}(P_{jk}+P_{j}a_{k}+\eta_{jk}P_{l}a_{l})-a_{j}(P_{ik}+P_{i}a_{k}+\eta_{ik}P_{l}a_{l})]\}+O\left( \lambda^{2}\right) ,  \nonumber \\
 \label{pijxk} \\
\left[ P_{ij},\hat{X}_{st}\right] &=&-i\left( \eta _{si}\eta _{tj}-\eta
_{ti}\eta _{sj}\right)\nonumber\\
&&+\frac{i\lambda }{2}[
\eta _{si}(P_{jt}+P_ja_t-(2c_1-1)P_ta_j)
-\eta_{ti}(P_{js}+P_{j}a_{s}-\left( 2c_{1}-1\right)P_{s}a_{j})\nonumber\\
&&-\eta _{sj}(P_{it}+P_{i}a_{t}-\left(2c_{1}-1\right)P_{t}a_{i})
+\eta _{tj}(P_{is}+P_{i}a_{s}-\left( 2c_{1}-1\right)P_{s}a_{i})]+O\left( \lambda ^{2}\right). \nonumber\\
&&
\label{pklxij}
\end{eqnarray}
{   We add that
 all of the generalized phase space relations, including the tensorial coordinates and momenta, depend on the $c_1$ parameter.
 In particular for $c_1=0$, we obtain   in place of (\ref{pjxi})
\begin{equation}\label{kappa_px_class}
\left[ P_{k},\hat{X}_{i}\right] =-i\eta_{ki}(1+{\lambda}a_jP_j)+i{\lambda}P_ia_k+{O(\lambda^2)}.
\end{equation}
Such commutator would correspond to the so-called classical basis of $\kappa$-Poincar\'e ({{named also the }}
 natural realization) \cite{Kresic-Juric:2007vgu}, \cite{Borowiec:2009vb}, \cite{diff_calc}.
 For $c_1=\frac{1}{2}$ we get:
\begin{eqnarray}
\label{kappa_px_weyl}
\left[ P_{k},\hat{X}_{i}\right]  &=&-i\eta_{ki}(1+\frac{1}{2}
{\lambda} a_jP_j)-\frac{i}{2}{\lambda}(P_{ik}-P_{k}a_i)+{O(\lambda^2)}
\end{eqnarray}
which is the result
corresponding to the so-called Weyl realization
 of $\kappa$-Poincar\'e algebra
  \cite{Kresic-Juric:2007vgu}.}\\

\textbf{i) Reduction to the $\kappa$-Minkowski and relation with the $\kappa$-de-Sitter case}

{When {$g_N=0$} the relations (\ref{alg_sog})
   describe the $\kappa $-Minkowski space-time with Lorentz covariance algebra as symmetry and 
   the cross commutators obtained above (\ref{pjxi})-(\ref{pklxij}) are reduced  to the particular case of $\kappa $-Minkowski phase space relations.   
Since the expressions provided in (\ref{pjxi}), (\ref{pjxis}) and (\ref{pklxij}) are $\beta$-independent they will remain the same in the reduction to $\kappa$-Minkowski case ($g_N=0$) up to the linear order in $\lambda$. The relation (\ref{pijxk}) reduces to:}
\begin{equation}
\left[ P_{ij},\hat{X}_{k}\right] =\frac{i
\lambda}{2}[\eta _{i}{}_{k}P_{jl}a_{l}+\left( 2c_{1}-1\right) \left(
P_{jk}a_{i}+P_{j}a_{k}a_{i}+\eta_{j}{}_{k}P_{l}a_{l}a_{i}\right) -(i\leftrightarrow j)]+{O(\lambda^2)}.
\end{equation}  
{
Basic quantum-deformed Heisenberg algebra relation is described by the cross commutation relation
  (\ref{pjxi}), which in  
  linear order of $\lambda$ does not depend on the $\beta$ parameter and  contains  only the terms
  generated by $\kappa$-deformation.}
 We know however from the standard Snyder model (see \cite{Snyder40})
 that the term linear in $\beta$ is bilinear in the momenta.
   Because terms
   bilinear in $P_i$
  are
   as well
   bilinear in $\lambda$, the parameter $\beta$ can contribute only  at {the}
     second
   perturbative order in $\lambda$.
   One can state 
   that for small $\lambda$
      the
   terms generated by $\kappa$-deformation dominate
    over the
   $\beta$-dependent terms, but  for more precise statement the {perturbative $\lambda^2$ order terms should also be calculated for the relation  (\ref{pjxi}).}\\
Further one can  compare
   these particular relations
    with the results obtained in the literature (see, e.g. in \cite{LN97}, \cite{TMPH}) where the Heisenberg double construction was
investigated for the $\kappa $-Minkowski
space-time and
quantum symmetry described by
 the $\kappa-$Poincar\'e algebra
(in Snyder model the quantum symmetry is linked with the de Sitter  algebra).\\

\textbf{ii) Reduction to the extended Snyder model}

We can reduce the 
 $\kappa$-dependent terms in commutation relations (\ref{pjxi})-(\ref{pklxij}) 
 and compare
  such results with  the  results 
 obtained from the Heisenberg double  of
  the extended Snyder
model \cite{MP_2021}.
 When
  $g_{0}= \ldots
  =g_{N-1}=0$ and
        $g_{N}=1$,
   the algebra (\ref{alg_sog}), after the change of variables (\ref{Xxrel}), (\ref{Pprel}), provides  the extended Snyder algebra.
 The phase space
  relation calculated  in
(\ref{pjxi})-(\ref{pklxij})  
 are reduced to  the following ones
\begin{eqnarray}\label{pkxi_ES}
\left[ p_{k},\hat{x}_{i}\right] &=&
-i\eta _{k}{}_{i}+i\lambda
c_{1}p_{ki}+O\left( \lambda ^{2}\right), \\
\left[ p_{k},\hat{x}_{ij}\right] &=&-i\lambda \left( 1-c_{1}\right) \left(
\eta _{ik}p_{j}-\eta _{jk}p_{i}\right) +O\left( \lambda ^{2}\right), \\
\left[ p_{ij},\hat{x}_{k}\right] &=&i\frac{\lambda }{2}\beta \left( \eta
_{i}{}_{k}p_{j}-\eta _{jk}p_{i}\right) +O\left( \lambda ^{2}\right), \\
\left[ p_{ij},\hat{x}_{st}\right] &=&
-i\left( \eta _{si}\eta _{tj}-\eta
_{ti}\eta _{sj}\right) +i\frac{\lambda }{2}[\left( \eta _{si}p_{jt}-\eta
_{sj}p_{it}\right) -\left( \eta _{ti}p_{js}-\eta _{tj}p_{is}\right)
]+O\left( \lambda ^{2}\right).\label{ES}
\end{eqnarray}%
 The cross commutation relations  in the generalized phase space obtained by the Heisenberg
double method  for 
  the extended Snyder model were calculated in \cite{MP_2021} and resulted in the following:
  \begin{eqnarray}
\left[ p_{k},\hat{x}_{i}\right] &=&
-i\eta _{k}{}_{i}+i\frac{\lambda}{2}p_{ki}+O\left( \lambda ^{2}\right), \\
\left[ p_{k},\hat{x}_{ij}\right] &=&-i\frac{\lambda}{2} \left(
\eta _{ik}p_{j}-\eta _{jk}p_{i}\right) +O\left( \lambda ^{2}\right), \\
\left[ p_{ij},\hat{x}_{k}\right] &=&i\frac{\lambda }{2}\beta \left( \eta
_{i}{}_{k}p_{j}-\eta _{jk}p_{i}\right) +O\left( \lambda ^{2}\right), \\
\left[ p_{ij},\hat{x}_{st}\right] &=&
-i\left( \eta _{si}\eta _{tj}-\eta
_{ti}\eta _{sj}\right) +i\frac{\lambda }{2}[\left( \eta _{si}p_{jt}-\eta
_{sj}p_{it}\right) -\left( \eta _{ti}p_{js}-\eta _{tj}p_{is}\right)
]+O\left( \lambda ^{2}\right).
\end{eqnarray}
They do agree with the above results (\ref{pkxi_ES})-(\ref{ES}) for $c_{1}=1/2$.

\section{Another Heisenberg double for the $\kappa$-extended Snyder model}

Using  the algebra (\ref{alg_sog}) describing $\kappa$-extended Snyder model, also
 other  Heisenberg double
  construction
  can be considered. If we introduce the algebra of functions generated by
   dual
    de Sitter group matrices $\Lambda _{\alpha \beta }$: $\{\Lambda _{\alpha \beta }:\left[ \Lambda
_{\alpha \beta },\Lambda _{\mu \nu }\right] =0:\Lambda ^{T}g\Lambda =g\}$ we
 should  postulate
\begin{eqnarray}
\Delta \left( \Lambda _{\rho \sigma }\right) &=&\Lambda _{\rho \alpha
}\otimes \Lambda _{\alpha \sigma };\quad \epsilon \left( \Lambda _{\rho
\sigma }\right) =g_{\rho \sigma }\quad ;S\left( \Lambda _{\rho \sigma
}\right) =(\Lambda ^{-1})_{\rho \sigma }=\Lambda _{\sigma \rho }.
\end{eqnarray}%
One can also introduce the matrices
$\tilde{\Lambda}_{\alpha \beta }$
which are related with the above group elements via 
 the map $\Lambda _{\rho \sigma
}=\left( O\tilde{\Lambda}O^{T}\right) _{\rho \sigma }$\footnote{
In this case we have: $\{\tilde{\Lambda}_{\alpha \beta
}:\left[ \tilde{\Lambda}_{\alpha \beta },\tilde{\Lambda}_{\mu \nu }\right]
=0:\tilde{\Lambda}^{T}\eta \tilde{\Lambda}=\eta \}$,
\begin{eqnarray}
\Delta \left( \tilde{\Lambda}_{\rho \sigma }\right) &=&\tilde{\Lambda}_{\rho
\alpha }\otimes \tilde{\Lambda}_{\alpha \sigma };\quad \epsilon \left(
\tilde{\Lambda}_{\rho \sigma }\right) =\eta _{\rho \sigma }\quad ;S\left(
\tilde{\Lambda}_{\rho \sigma }\right) =(\tilde{\Lambda}^{-1})_{\rho \sigma }=%
\tilde{\Lambda}_{\sigma \rho }.\nonumber
\end{eqnarray}%
Such algebra was also used in
\cite{MP_2021} in our studies of the Heisenberg
double for the extended Snyder model.},
 ($\alpha ,\beta =0,1,\ldots ,N$). The
basic duality relation is given by:
\begin{equation}
<\Lambda _{\rho \sigma },\hat{X}_{\mu \nu }>=-i\lambda (g_{\rho \mu
}g_{\sigma \nu }-g_{\rho \nu }g_{\sigma \mu }).  \label{dual}
\end{equation}%
We consider the
 following
 left Hopf action $\triangleright $:
\begin{equation}
\Lambda _{\rho \sigma }\triangleright \hat{X}_{\mu \nu }=<\Lambda _{\rho
\sigma },\hat{X}_{\mu \nu _{\left( 2\right) }}>\hat{X}_{\mu \nu _{\left(
1\right) }}=g_{\rho \sigma }\hat{X}_{\mu \nu }-i\lambda (g_{\rho \mu
}g_{\sigma \nu }-g_{\rho \nu }g_{\sigma \mu })
\end{equation}%
and we obtain the cross commutation relations
 which are defined by the Heisenberg
double  method
\begin{eqnarray}
\label{45}
\left[ \Lambda _{\rho \sigma },\hat{X}_{\mu \nu }\right] &=&\hat{X}_{\mu \nu
}{}_{\left( 1\right) }<\Lambda _{\rho \sigma \left( 1\right) },\hat{X}_{\mu
\nu \left( 2\right) }>\Lambda _{\rho \sigma \left( 2\right) }-\hat{X}_{\mu
\nu }\Lambda _{\rho \sigma }  \label{Lx} \\
&=&-i\lambda (g_{\rho \mu }\Lambda _{\nu \sigma }-g_{\rho \nu }\Lambda _{\mu
\sigma }).\nonumber
\end{eqnarray}
If we recall the relations
$\hat{X}_{kN}=\kappa \hat{X}_{k}$
 and
 $g_{ij}=\eta _{ij},\quad g_{iN}=\kappa a_{i},\quad
g_{NN}=g_{N}=\kappa ^{2}\beta $,
 we get from (\ref{45})
  the cross commutation relations between the dual group
elements and quantum algebra generators
\begin{eqnarray}
\left[ \Lambda _{jk},\hat{X}_{i}\right] &=&-\frac{i\lambda }{\kappa }%
(g_{ji}\Lambda _{Nk}-g_{jN}\Lambda _{ik})=-\frac{i\lambda }{\kappa }(\eta
_{ji}\Lambda _{Nk}-\kappa a_{j}\Lambda _{ik}),  \label{LX1} \\
\left[ \Lambda _{jN},\hat{X}_{i}\right] &=&-\frac{i\lambda }{\kappa }%
(g_{ji}\Lambda _{NN}-g_{jN}\Lambda _{iN})=-\frac{i\lambda }{\kappa }(\eta
_{ji}\Lambda _{NN}-\kappa a_{j}\Lambda _{iN}),  \label{LX2} \\
\left[ \Lambda _{Nk},\hat{X}_{i}\right] &=&-i\lambda (a_{i}\Lambda
_{Nk}-\kappa \beta \Lambda _{ik}),  \label{LX3} \\
\left[ \Lambda _{NN},\hat{X}_{i}\right] &=&-i\lambda (a_{i}\Lambda
_{NN}-\kappa \beta \Lambda _{iN}).  \label{LX4}
\end{eqnarray}%
For  the tensorial coordinates $\hat{X}_{ij}$
  we obtain the cross relations:
\begin{eqnarray}
\left[ {\Lambda}_{lk},\hat{X}_{ij}\right] &=&-i\lambda (\eta _{li}\Lambda
_{jk}-\eta _{lj}\Lambda _{ik}),  \label{LX5} \\
\left[{\Lambda}_{lN},\hat{X}_{ij}\right] &=&-i\lambda (\eta _{li}\Lambda
_{jN}-\eta _{lj}\Lambda _{iN}),  \label{LX6} \\
\left[ {\Lambda}_{Nk},\hat{X}_{ij}\right] &=&-i\lambda \kappa (a_{i}\Lambda
_{jk}-a_{j}\Lambda _{ik}),  \label{LX7} \\
\left[ {\Lambda}_{NN},\hat{X}_{ij}\right] &=&-i\lambda \kappa (a_{i}\Lambda
_{jN}-a_{j}\Lambda _{iN}).  \label{LX8}
\end{eqnarray}
Note that $\Lambda _{ij}=\tilde{\Lambda}_{ij}$\footnote{%
Footnote {${}^{5)}$} can be supplemented by the following relations obtained
from (46), (50):
\begin{eqnarray}
\left[ \tilde{\Lambda}_{jk},\hat{X}_{i}\right] &=&-\frac{i\lambda }{\kappa }
(\eta _{ji}\tilde{\Lambda} _{Nk}-\kappa a_{j}\tilde{\Lambda} _{ik}), \\
\left[ \tilde{\Lambda}_{lk},\hat{X}_{ij}\right] &=&-i\lambda (\eta
_{li}\tilde{\Lambda} _{jk}-\eta _{lj}\tilde{\Lambda} _{ik}).
\end{eqnarray}%
}. 
One can check explicitly that the duality $<\Lambda _{\rho \sigma },\hat{X}%
_{\mu \nu }>$ given above reduces to
\begin{equation}
<\tilde{\Lambda}_{\rho \sigma },\hat{x}_{\mu \nu }>=-i\lambda (\eta _{\rho
\mu }\eta _{\sigma \nu }-\eta _{\rho \nu }\eta _{\sigma \mu })
\end{equation}%
for the extended Snyder model (i.e. when $g_{Ni}=g_{0}=...=g_{N-1}=0,g_{ij}=\eta
_{ij},g_{NN}=g_{N}=\kappa ^{2}\beta $).
 Therefore, all the cross commutation relations  (\ref{LX1})-(\ref{LX8})
 are the same as  the ones  obtained in \cite{MP_2021}
where the extended Snyder model was investigated.

In the reduction of  the above results to the $\kappa$-Minkowski case ($g_N = 0$)
 only two relations { (\ref{LX3}), (\ref{LX4})} depend on $\beta$ and  we easily  see  the  result of reduction.
Further, focusing on
 the relation  (\ref{LX1}),
 after putting
  $a_{j}=\frac{1}{\kappa }\delta _{j}^{0}$ (i.e. assuming a time-like
$\kappa-$deformation)
we obtain the following covariance relations
\begin{equation}
\left[ {\Lambda }_{jk},\hat{X}_{i}\right] =
-\frac{i}{\kappa }{\lambda}\left(\eta _{ji}\Lambda _{Nk}-\delta _{j}^{0}\Lambda _{ik} \right).
\end{equation}

\section{Discussion and outlook}
 For
  the $\kappa$-extended Snyder model introduced in \cite{MM_2021} we
    constructed  the generalized quantum phase space, which depends on three parameters, $\beta$, $\kappa$ and $\lambda$.
    The Hopf algebra describing the quantum symmetries of the model is coassociative, hence one can use the Heisenberg double construction for the correct description of the respective generalized quantum phase space.
      Since {this} model contains  Snyder
      ($\beta \neq 0$) and $\kappa$-Minkowski sectors
      ($\kappa \neq 0)$, one can discuss how the generalized quantum phase space can be reduced to these special cases.
 The   algebraic    relations  determining
    the generalized phase space
          for the $\kappa $-extended Snyder model  have been calculated to
{the} first order in the $\lambda$ parameter, and for most commutation relations to such order  we get only  the formulae
       modified by  $\kappa$-deformation. Our algebra also includes the orthogonal (de Sitter) algebra generators, which are described by tensorial coordinates.

The quantum group
  describing quantum symmetries  of  the $\kappa$-extended Snyder model is built up from deformed coproducts of momenta \cite{MM_2021Sept} and undeformed Lorentz
  algebra relations.
  {When $\kappa\to\infty$ the $\kappa$-extended Snyder model reduces to the extended Snyder model which describes the noncommutative de-Sitter space-time, with $\beta$ as the inverse square of constant curvature parameter $R$ characterizing dual de-Sitter pseudosphere in momentum space. Considering the case, when $\beta\to 0$ but keeping parameter $\kappa$ we obtain the known $\kappa$-deformed Minkowski spacetime.
  Keeping both parameters, one gets an analogue of $\kappa$-deformation of noncommutative de-Sitter space related with the classical r-matrices of de-Sitter algebra \cite{Bal1, Bal2}. Such classical r-matrices, which introduce two de-Sitter parameters, namely one related with de-Sitter curvature $R$ and other related with $\kappa$ deformation, can be considered and it would be interesting to investigate the connection between $\kappa$-deformed extended Snyder model and the deformations introducing the $\kappa$-deformations of de-Sitter geometry. 
} {Also, it would be of interest to see what realizations of  $\kappa$-Poincar\'e,
    after the  interpretation of  $\beta$ as the    inverse square of de Sitter radius
     ($\beta \sim R^{-2}$), could
   be obtained  by the quantum Inonu-Wigner contraction procedure.}\\
 The Heisenberg double
  for the  $\kappa$-Poincar\'e quantum group leads to the $\kappa -$deformed phase space
   that was investigated in  different bases for the $\kappa $-Poincar\'e Hopf algebra
 \cite{KosMas}-
\cite{TMPH}, 
  \cite{MM_2021Sept}. 
In $\kappa$-extended Snyder
  model  with different quantum symmetry group and
 different coalgebra sector,
 the  comparison between these two cases and the
   present model
   involving tensorial coordinates should still be discussed.
%


Further, one of the tasks
which would be interesting to investigate is the Hopf algebroid  providing
 the generalized quantum phase
spaces of the $\kappa$-extended Snyder model. Hopf algebroids
describing the quantum phase spaces with NC  space-time coordinates have been
studied  in the literature recently \cite{Halg2},
\cite{Juric:2013mma}-
 \cite{Lukierski:2018ulb},
 and it should be recalled
  that
   Heisenberg double construction provides a natural example of the Hopf
bialgebroid structure \cite{JHL}.



%
%
%
%
%

It can finally be added that the parameters $\beta$, $\kappa$ and $\lambda$
  are not
treated here as genuine quantum deformations  parameters
 introducing quantum deformation determined by classical
$r$-matrices, which
satisfy classical or modified Yang Baxter equations.
{These parameters appear 
in the procedure of changing the ${o}({1,N})$ basis, which effectively leads to the modified de Sitter relations
 and noncanonical modified  de Sitter metric $g_ {\mu\nu}$.}
   The
  parameters $\beta, \kappa$ and $\lambda$
appear as determining the algebra basis and
  in this way  we
 introduce physical parameters ($\beta$ is related with de Sitter radius, $\kappa$ usually
 is linked with Planck mass, and {the parameter $\lambda$ is related with the Planck constant}).
 Thus, in
 order to describe the
   $\kappa$-extended Snyder model we only redefine the
 standard basis of classical de Sitter algebra, this mathematically rather trivial
 operation leads to results which
 might be
  significant in physical applications.

\section*{Acknowledgements}

S. Meljanac, S. Mignemi and A. Pacho\l \ \
 would like to acknowledge the support of the European Cooperation in Science and Technology COST Action CA18108. S. Mignemi acknowledges support from GNFM. J.~Lukierski and A.  Pacho\l ~were supported by Polish  NCN grant 2017/27/B/ST2/01902; J. Lukierski would
like also to thank A. Borowiec and M. Woronowicz for valuable comments.

\end{document}